\documentclass[12pt]{article}
\usepackage{amsfonts,amssymb,amsmath}
\usepackage[T1]{fontenc}
\usepackage{authblk}

\usepackage{graphicx}

\usepackage{tikz}
\usetikzlibrary{arrows,snakes,backgrounds}
\usepackage{subcaption}

\newcommand{\ZZ}{{\mathbb Z}}

\newcommand{\ra}{\rightarrow}
\newcommand{\eps}{\epsilon}
\newcommand{\g}{\gamma}
\newcommand{\ket}[1]{{\vert {#1}\rangle}}

\def\~#1{\widetilde{#1}}
\def\_#1{_{\mathrm{#1}}}

\def\bel#1{\begin{equation}#1\end{equation}}

\title{Exact bosonization in two spatial dimensions and a new class of lattice gauge theories}
\author[*]{Yu-An Chen} 
\author[*]{Anton Kapustin}
\author[**]{\DJ or\dj e Radi\v cevi\'c}
\affil[*]{California Institute of Technology, Pasadena, CA 91125, USA}
\affil[**]{Perimeter Institute for Theoretical Physics, Waterloo, Ontario, Canada N2L 2Y5}

\begin{document}

\maketitle
\begin{abstract}
We describe a 2d analog of the Jordan-Wigner transformation which maps an arbitrary fermionic system on a 2d lattice to a lattice gauge theory while preserving the locality of the Hamiltonian. When the space is simply-connected, this bosonization map is an equivalence. We describe several examples of 2d bosonization, including free fermions on square and honeycomb  lattices and the Hubbard model. We describe Euclidean actions for the corresponding lattice gauge theories and find that they contain Chern-Simons-like terms. Finally, we write down a fermionic dual of the gauged Ising model (the Fradkin-Shenker model).

\end{abstract}

\section{Introduction}

It is well known that the Jordan-Wigner transformation establishes an equivalence between the quantum Ising chain in a transverse magnetic field and a system of free spinless fermions on a 1d lattice. This equivalence is very useful and is the quickest way to solve the 2D Ising model.\footnote{2D here means ``two Euclidean space-time dimensions.''}

The use of the Jordan-Wigner transformation is not limited to the quantum Ising chain. It establishes a very general kinematic equivalence between 1d fermionic systems and 1d spin chains with a $\ZZ_2$ spin symmetry. One can regard it as a very special isomorphism between the algebra of fermionic observables with trivial fermion parity and the $\ZZ_2$-even subalgebra of the algebra of observables in a spin chain. Its distinguishing feature is that it maps local observables\footnote{That is, observables that act nontrivially on a finite number of lattice sites.} with trivial fermion parity on the fermionic side to local observables that commute with the total spin parity
$$
(-1)^{S^z}=\prod_i (-1)^{S^z_i}
$$
on the bosonic side. Thus any local Hamiltonian for a 1d fermionic system can be mapped to a local spin chain Hamiltonian which preserves $S^z$ modulo $2$. In this sense, the Jordan-Wigner transformation is local.

There are several ways to generalize the Jordan-Wigner transformation to 2d lattice systems. One obvious approach is to take a square lattice, represent it as a 1d system by picking a lattice path which snakes through the whole lattice and  visits each site once, and apply the 1d Jordan-Wigner transformation. This leads to a bosonization map which maps some, but not all, local observables with a trivial fermion parity to local observables with trivial $S^z$. The lack of 2d locality causes problems, since even very simple fermionic Hamiltonians are mapped to non-local spin Hamiltonians and vice versa. But there are interesting exceptions \cite{JWcompass}, the Kitaev honeycomb model \cite{Kitaevanyons} being among them \cite{honeycombone, honeycombtwo, honeycombthree}. Another  approach to 2d bosonization is to use flux attachment \cite{Fradkin1989, WangA}: a fermion is represented by a boson interacting with a Chern-Simons gauge field. Related ideas in the continuum have been the focus of much recent interest \cite{guy, shiraz, son, radicevic, ofer, karchtong, murugan, ssww, hsin, rtt, kachru, raghu, komargodski}. However, it is hard to make this precise on the lattice, due to well-known difficulties with defining a lattice Chern-Simons theory. A popular strategy is to eliminate the Chern-Simons gauge field by solving its equations of motion, but this again leads to a non-local map. 

A very interesting example of exact 2d bosonization, or rather fermionization, was presented by A.\ Kitaev in his paper on the honeycomb model \cite{Kitaevanyons}. At first it appears quite special, but in fact it provides a method for mapping an arbitrary system of Majorana fermions on a trivalent lattice to a system of bosonic spins on the same lattice. The spin Hilbert space is not a tensor product over all sites, but rather a subspace defined by a set of commuting constraints. There is one  constraint for each face of the lattice, indicating that the dual bosonic system is a gauge theory. But it is a very unusual gauge theory, since the gauge field is a composite of spins. 

The goal of this paper is to describe a 2d  analog of the Jordan-Wigner transformation  which obeys locality, and to give some examples of 2d bosonization. We will show that any 2d fermionic system on a lattice can be mapped to a system of bosons so that, on a topologically trivial space, this map is an equivalence and every local fermionic Hamiltonian maps to a local spin Hamiltonian. The main novelty compared to the 1d case is that the bosonic system is a $\ZZ_2$ gauge theory. This means that the Hilbert space is not a tensor product of local Hilbert spaces, but a subspace in such a tensor product defined by a set of commuting local constraints. They can be interpreted as Gauss law constraints. 

Our bosonization procedure shares some similarities both with the flux-attachment approach and with Kitaev's approach. It follows the same strategy as the flux-attachment approach but uses a lattice $\ZZ_2$ gauge field in place of a $U(1)$ Chern-Simons gauge field. There is no problem writing down a Chern-Simons-like term for a $\ZZ_2$ gauge field. An additional  benefit is that we do not need to introduce separate bosonic degrees of freedom to which the flux is attached: we make use only of degrees of freedom that are already present in the gauge field. Our bosonization procedure is completely general and local, just like in Kitaev's approach, but there are couple of differences too: (1) the fermions are complex rather than Majorana, so that the fermionic Hilbert space is naturally a tensor product over all sites; (2) the bosonic variables live on edges rather than on sites, and the gauge field is fundamental rather than composite.

The connection between gauge symmetry and bosonization rests on the observation made in \cite{GaiottoAK} that 2d bosonization should map fermionic systems to bosonic systems with a global 1-form $\ZZ_2$ symmetry and a suitable 't Hooft anomaly. On a lattice, global 1-form symmetries can exist only in gauge theories. The proposal of \cite{GaiottoAK} was made concrete in \cite{BGK} for topological systems (that is, spin-TQFTs), but it was implicit in that paper that the same strategy should apply for general fermionic systems on a lattice. In this paper we make this completely explicit. Namely, we show that on a simply connected space one can isomorphically map the bosonic subalgebra of the algebra of local fermionic observables to the algebra of local gauge-invariant observables in a suitable $\ZZ_2$ lattice gauge theory.

The bosonization map is not canonical and depends on some additional choices which depend on the type of lattice. We discuss two kinds of lattices: the square lattice and an arbitrary triangulation endowed with a branching structure. In the latter case the fermions live on the faces of the triangulation.

Gauss law constraints for gauge theories dual to 
fermionic systems are not standard. Their meaning becomes clearer if we discretize time and consider the corresponding Euclidean lattice partition function. It turns out that the unusual Gauss law arise from a Chern-Simons-like term in the action. These terms necessarily break invariance under the cubic symmetry (if one starts from a 2d square lattice). 

The paper is organized as follows. In Section 2 we construct the 2d bosonization map on the kinematic level, i.e.\ on the level of the algebras of observables. In Section 3 we give several examples of bosonization for concrete fermionic systems, such as free fermions and the Hubbard model. Conversely, we describe the fermionization of the simplest lattice gauge theories with a non-standard Gauss law. Such gauge theories are dual to free fermionic theories and thus are integrable. In Section 4 we derive the Euclidean partition function for the gauge theories considered in Section 3. In Section 5 we show that the gauged 3D Ising model, also known as the Fradkin-Shenker model \cite{FS}, has a fermionic description involving a system of fermions coupled to a $\ZZ_2$ gauge field. This illustrates that the 2d analog Jordan-Wigner transformation applies also to fermionic systems with gauge symmetries. In one of the Appendices we make a detailed comparison between our bosonization procedure and that proposed by A. Kitaev \cite{Kitaevanyons}.

A.\ K.\ would like to thank T.\ Senthil for a discussion. 
The research of A.\ K.\ and Y.\ C.\ was supported in part by the U.S.\ Department of Energy, Office of Science, Office of High Energy Physics, under Award Number DE-SC0011632. A.\ K.\ was also supported by the Simons Investigator Award.  Research at Perimeter Institute is supported by the Government of Canada through Industry Canada and by the Province of Ontario through the Ministry of Economic Development \& Innovation.

\section{Bosonization on a two-dimensional lattice}

\subsection{Square lattice}\label{sec:square}

We first introduce our bosonization method on an infinite 2d square lattice. Suppose that we have  a model with fermions living at the centers of faces. Let us describe the generators and relations in the algebra of local observables with trivial fermion parity (the even fermionic algebra for short).

On each face $f$ we have a single fermionic creation operator $c_f$ and a single fermionic annihilation operator $c^\dagger_f$ with the usual anticommutation relations. The fermionic parity operator on face $f$ is $(-1)^{F_f}=(-1)^{c_f^\dagger c_f}$. It is a ``$\ZZ_2$ operator'' (i.e.\ it squares to $1$). All operators $(-1)^{F_f}$ commute with each other. The even fermionic algebra is generated by these operators and the operators $c_f^\dagger c_{f'}$, $c_f c_{f'}$, and their Hermitean conjugates, where $f$ and $f'$ are two faces which share an edge. Overall, we get four generators for each edge and one generator for each face. 

In fact, one can make do with a single generator for each edge and a single generator for each face, provided we choose a consistent orientation of all faces and arbitrary  orientations of all edges. Following \cite{BGK}, we introduce Majorana fermions
$$
\gamma_f=c_f+c_f^\dagger,\quad \gamma'_f=(c_f-c_f^\dagger)/i.
$$
Then the operators 
$$
(-1)^{F_f}=-i\gamma_f\gamma'_f, 
$$
and
$$
S_e=i\gamma_{L(e)}\gamma'_{R(e)},
$$
are $\ZZ_2$ operators and generate the even algebra. Here $L(e)$ and $R(e)$ are faces to the left and to the right of the edge $e$ with respect to the chosen orientations.\footnote{That is, $L(e)$ is the face which induces the same orientation on $e$ as the given orientation of $e$, while $R(e)$ is the face which induces the opposite orientation.} We will refer to $S_e$ as the hopping operator for edge $e$. It anticommutes with $(-1)^{F_f}$ if $f=L(e)$ or $f=R(e)$ and commutes with all other $(-1)^{F_f}$. 

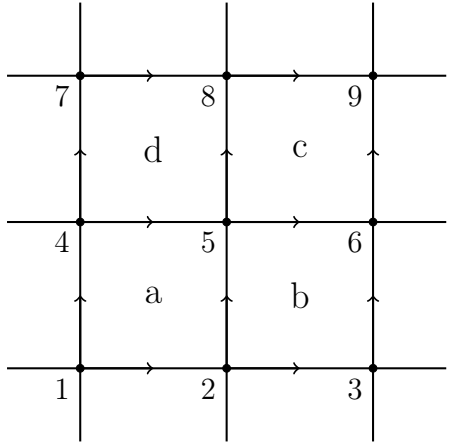
\begin{figure}
\centering
\resizebox{6cm}{!}{%
\begin{tikzpicture}
\draw[thick] (-3,0) -- (3,0);\draw[thick] (-3,-2) -- (3,-2);\draw[thick] (-3,2) -- (3,2);
\draw[thick] (0,-3) -- (0,3);\draw[thick] (-2,-3) -- (-2,3);\draw[thick] (2,-3) -- (2,3);
\draw[->] [thick](0,0) -- (1,0);\draw[->][thick] (0,2) -- (1,2);\draw[->][thick] (0,-2) -- (1,-2);
\draw[->][thick] (0,0) -- (0,1);\draw[->][thick] (2,0) -- (2,1);\draw[->][thick](-2,0) -- (-2,1);
\draw[->][thick] (-2,0) -- (-1,0);\draw[->][thick] (-2,2) -- (-1,2);\draw[->][thick](-2,-2) -- (-1,-2);
\draw[->][thick] (-2,-2) -- (-2,-1);\draw[->][thick] (0,-2) -- (0,-1);\draw[->] [thick](2,-2) -- (2,-1);
\filldraw [black] (-2,-2) circle (1.5pt) node[anchor=north east] {1};
\filldraw [black] (0,-2) circle (1.5pt) node[anchor=north east] {2};
\filldraw [black] (2,-2) circle (1.5pt) node[anchor=north east] {3};
\filldraw [black] (-2,-0) circle (1.5pt) node[anchor=north east] {4};
\filldraw [black] (0,0) circle (1.5pt) node[anchor=north east] {5};
\filldraw [black] (2,0) circle (1.5pt) node[anchor=north east] {6};
\filldraw [black] (-2,2) circle (1.5pt) node[anchor=north east] {7};
\filldraw [black] (0,2) circle (1.5pt) node[anchor=north east] {8};
\filldraw [black] (2,2) circle (1.5pt) node[anchor=north east] {9};
\draw (-1,-1) node{\large a};
\draw (1,-1) node{\large b};
\draw (1,1) node{\large c};
\draw (-1,1) node{\large d};
\end{tikzpicture}
}
\caption{Bosonization on a square lattice requires constraints on vertices.}
\label{fig:square}
\end{figure}

Other relations depend on the choice of orientations. We will choose the usual (counterclockwise) orientation of the plane and point all horizontal edges to the east, and all vertical edges to the north; see Fig.~\ref{fig:square}. Then it is easy to see that $S_e$ and $S_{e'}$ may fail to commute only if $e$ and $e'$ share a point and are perpendicular. If $e$ and $e'$ share a point and are perpendicular, then in the notation of 
Fig.~\ref{fig:square} we have
\begin{equation}
\left[S_{56},S_{58}\right]=\left[S_{25},S_{45}\right]=0,\quad \left\{S_{25},S_{56}\right\}=\left\{S_{58},S_{45}\right\}=0.
\end{equation}
In other words, $S_e$ and $S_{e'}$ anticommute if $e$ and $e'$ issue from the same vertex and point either east and south, or north and west. They commute in all other cases.

Additional relations emerge if we consider the product of four hopping operators corresponding to all edges issuing from a vertex. This corresponds to an operator taking a fermion full circle around the vertex. The resulting operator commutes with $(-1)^{F_f}$ for all $f$ and thus must be some function of these operators. Indeed, a short calculation shows that
\begin{equation}
\begin{split}
 S_{{58}} S_{{56}} S_{{25}} S_{{45}} &= (i \g_d \g^\prime_c)  (i \g_c \g^\prime_b)  (i \g_a \g^\prime_b)  (i \g_d \g^\prime_a) \\
&= (i \g^\prime_a \g_a) (i \g^\prime_c \g_c) \\
&= (-1)^{F_a} (-1)^{F_c}.
\end{split}
\label{eq: S delta square}
\end{equation}
It is clear intuitively and can be shown rigorously (see Appendix for a sketch of a proof) that these are all relations between our chosen generators if the lattice is infinite, or if it is finite but topologically trivial. 

The dual description will consist of bosonic spins living on edges of the same lattice. The operators acting on each edge $e$ are Pauli matrices $\sigma^x_e$, $\sigma^y_e$, and $\sigma^z_e$. To reduce notation clutter, we denote them $X_e$, $Y_e$, and $Z_e$. This is the usual operator algebra of the toric code.

We have two kinds of edges: edges oriented  east and edges oriented north. If $e$ is oriented east (resp.\ north), let $r(e)$ be the edge which points north (resp.\ east) and  ends where $e$ begins. In the notation of Fig.~\ref{fig:square}, $r(e_{56})=e_{25}$, $r(e_{58})=e_{45}$. 
It will be useful to define the composite operator
\begin{equation}
U_e = X_e Z_{r(e)}.
\end{equation}
In the toric code language, $U_e$ is the operator moving the $\eps$-particle across edge $e$.
We also define the ``flux operator'' at each face $f$ to be
\begin{equation}
W_f = \prod_{e \subset f} Z_e.
\end{equation}

Our bosonization map is defined as follows:
\begin{enumerate}

\item
We identify the fermionic states $ |F_f=0 \rangle$ and $ |F_f=1 \rangle$  with bosonic states for which $W_f=1$ and $W_f=-1$, respectively. This amounts to dualizing
\begin{equation}
(-1)^{F_f} = -i \g_f \g^\prime_f \longleftrightarrow W_f.
 \label{eq:bosonization assumption 1}
\end{equation}

\item
The fermionic hopping operator $S_e$ is identified with $U_e$ defined above,
\begin{equation}
 S_e =  i \g_{L(e)} \g^\prime_{R(e)} \longleftrightarrow U_e.
 \label{eq:bosonization assumption 2}
\end{equation}
\end{enumerate}
All operator relations discussed above are preserved under this map. The only exception is the relation (\ref{eq: S delta square}), which is absent on the bosonic side. Instead, the product $S_{58} S_{56} S_{25} S_{45}$ maps to
\begin{equation}
U_{{58}} U_{{56}} U_{{25}} U_{{45}} =  W_{f_a} \prod_{e \supset v_5} X_e.
\label{eq: U delta square}
\end{equation}
To get an algebra homomorphism, we must impose a constraint on the bosonic variables at vertex $5$, namely
\begin{equation}
W_{f_c} \prod_{e \supset v_5} X_e  =1.
\end{equation}
For a general vertex $v$, the constraint is 
\begin{equation}
W_{\text{NE}(v)} \prod_{e \supset v} X_e = 1.
\label{eq:gauge constraint at vertex}
\end{equation}
where $\text{NE}(v)$ is the face northeast of $v$.

We interpret this as a Gauss law for the bosonic system. The presence of the Gauss law means that we are dealing with a gauge theory. Since the constraint at each vertex  is a $\ZZ_2$ operator, this is a $\ZZ_2$ gauge theory. The algebra of gauge-invariant observables on the bosonic side (i.e.\ the algebra of local observables commuting with all Gauss law constraints) is generated by operators $U_e$ and $W_f$, and there are no further relations between them apart from those which exist between $S_e$ and $(-1)^{F_f}$. Thus the above map is an isomorphism and defines a 2d version of the Jordan-Wigner transformation. 

The constraint \eqref{eq:gauge constraint at vertex} couples the electric charge at a vertex $v$ to the magnetic flux at face $\text{NE}(v)$. Thus our modified Gauss law implements charge-flux attachment, and it is not surprising that operators $U_e$ which move the flux behave as fermionic bilinears.

Note also that the total fermion number operator\footnote{Not to be confused with the fermion parity operator $\prod_f (-1)^{F_f}$.}
$$
F=\sum_f \frac12(1+i\gamma_f\gamma'_f)
$$
is mapped to the net magnetic flux
$$
\sum_f \frac12 (1-W_f).
$$
While the fermion number operator is ultra-local (it is a sum of operators each of which acts nontrivially only on fermions at a particular site), its bosonized version is not ultra-local.

\subsection{Triangulation}\label{sec:triang}

The bosonization method described above also works for any triangulation $T$ with a branching structure.\footnote{A branching structure on a triangulation is an orientation for every edge such that for every face the oriented edges do not form an oriented loop. A  branching structure specifies an ordering of vertices of every face: on each face $f$ there will be exactly one vertex (denoted $f_0$) with two edges of $f$ oriented away from the vertex, one vertex ($f_1$) with one edge of $f$ entering it and one leaving it, and another vertex ($f_2$) with two edges of $f$ oriented towards it.} The main idea of this approach was previously described in \cite{BGK}. We will review this material first and then describe a general way to perform bosonization on a 2d triangulation.

We assume again that we are given a global orientation and for an edge $e$ define $L(e)$ and $R(e)$ to be the faces to the left and to the right of $e$, just as for the square lattice. On a face $f$ we have fermionic operators $a_f,a_f^\dagger$, or equivalently a pair of Majorana fermions $\gamma_f,\gamma'_f$. They are generators of a Clifford algebra. The fermion parity on face $f$ is $(-1)^{F_f}=-i\gamma_f\gamma'_f$.
A $\ZZ_2$ fermionic hopping operator on an edge $e$ can again be defined by $S_e = i \g_{L(e)} \g^\prime_{R(e)}$.

The even fermionic algebra is generated by $(-1)^{F_f}$ and $S_e$ for all faces and edges. The relations between them can also be described. Obviously, these operators are $\ZZ_2$, and $S_e$ anticommutes with $(-1)^{F_f}$ whenever $e\subset f$ and commutes with it otherwise. The operators $S_e$ and $S_{e'}$ sometimes commute and sometimes anticommute. To describe the commutation rule more precisely, it is convenient to use the cup product on mod-2 1-cochains. Recall that a mod-2 $p$-cochain is a $\ZZ_2$-valued function on $p$-simplices of the triangulation. In our case, $p$ can be $0$, $1$, or $2$, corresponding to functions on vertices, edges, and faces respectively. The cup product of two 1-cochains is a 2-cochain defined as follows:
$$
(\alpha\cup\beta)(f_0,f_1,f_2)=\alpha(f_0, f_1) \beta(f_1, f_2).
$$
Here $\alpha$ and $\beta$ are arbitrary 1-cochains with values in $\ZZ_2$, and $f_0$, $f_1$, and $f_2$ are vertices of the face $f$, ordered in accordance with the branching structure. The cup product is not commutative (or supercommutative, which is the same thing since we are working modulo $2$). Let $\delta_e$ be the 1-cochain which takes value $1$ on the edge $e$ and value $0$ on all other edges. Then the commutation rule for $S_e$ and $S_e'$ is
\begin{equation}
S_e S_{e^\prime}= (-1)^{\int \delta_e \cup \delta_{e^\prime} + \delta_{e^\prime} \cup \delta_e} S_{e^\prime} S_e.
\label{eq:Se commutation}
\end{equation}
Here the integral of a 2-cochain is simply the sum of its values on all faces. In other words, if $e$ and $e'$ are distinct edges, $S_e$ and $S_{e'}$ anticommute if $e$ and $e'$ belong the same face and their union does not contain the edge $f_{02} = (f_0, f_2)$ of that face. They commute otherwise. 

There is also a relation for each vertex $v$, analogous to \eqref{eq: S delta square}, which reads
\begin{equation}\label{eq:vertexrel}
\prod_{e\supset v} S_e=c(v) \prod_{f\supset I^{02}_v} (-1)^{F_f},
\end{equation}
where $I^{02}_v$ is the set of those faces $f$ for which $v$ is either $f_0$ or $f_2$, and $c(v)$ is a certain $c$-number sign which depends on the vertex $v$. Its explicit form is not important for our purposes and can be found  in \cite{BGK}. 

To reproduce these relations in a bosonic model, we again introduce a spin variable for every edge and let $X_e,Y_e,Z_e$ be the corresponding Pauli matrices. We let
\begin{equation}\label{eq:Zf}
W_f=\prod_{e\subset f} Z_e,
\end{equation}
as before. This operator measures flux through face $f$. We anticipate that the bosonic model will be a gauge theory, and thus the algebra of gauge-invariant observables will be generated by $W_f$ and $\ZZ_2$ operators $U_e$ which anticommute with $W_f$ if $e\subset f$ and commute with $W_f$ otherwise. We also anticipate that in order for $U_e$ to behave as fermion hopping operators, we must implement charge-flux attachment. Our convention will be that if $W_f=-1$ for some face $f$, then electric charge will be sitting at the vertex $f_0$ of $f$. Then the flux hopping operator will take the form
\begin{equation}
U_e = X_e \prod_{f\in\{L(e),\, R(e)\}} Z_{f_{01}}^{\delta_e (f_{12})},
\label{eq:Ue definition}
\end{equation}
where $f_{ij}$ denotes the edge of $f$ connecting vertices $f_i$ and $f_j$. 

\begin{figure}
\centering
\begin{tikzpicture}[scale=1]
\draw[thick] (0,0) -- (-2,0);\draw[thick] (0,0) -- (-1,1.5);\draw[thick] (0,0) -- (-0.5,-2);
\draw[thick] (-2,0) -- (-0.5,-2);\draw[thick] (-2,0) -- (-1,1.5);\draw[thick] (0,0) -- (1.5,1);
\draw[thick] (-1,1.5) -- (1.5,1);\draw[thick] (0,0) -- (1,-1.5);\draw[thick] (-0.5,-2) -- (1,-1.5);
\draw[thick] (1.5,1) -- (1,-1.5);\draw[thick] (3,0) -- (1.5,1);
\draw[thick] (3,0) -- (1,-1.5);
\draw[->][thick] (0,0) -- (-1,0);\draw[->][thick] (-1,1.5) -- (-0.5,0.75);
\draw[->][thick] (0,0) -- (0.75,0.5);\draw[->][thick] (-0.5,-2) -- (-0.25,-1);
\draw[->][thick] (-0.5,-2) -- (-1.25,-1);\draw[->][thick] (-0.5,-2) -- (0.25,-1.75);
\draw[->][thick] (1,-1.5) -- (0.5,-0.75);\draw[->][thick] (1,-1.5) -- (1.25,-0.25);
\draw[->][thick] (1,-1.5) -- (2,-0.75);\draw[->][thick] (-1,1.5) -- (-1.5,0.75);
\draw[->][thick] (-1,1.5) -- (0.25,1.25);\draw[->][thick] (1.5,1) -- (2.25,0.5);
\filldraw [black] (-0.5,-2) circle (1.25pt) node[anchor=north] {0};
\filldraw [black] (1,-1.5) circle (1.25pt) node[anchor=north] {1};
\filldraw [black] (-1,1.5) circle (1.25pt) node[anchor=south] {2};
\filldraw [black] (0,0) circle (1.25pt) node[anchor=south] {3};
\filldraw [black] (-2,0) circle (1.25pt) node[anchor=east] {4};
\filldraw [black] (1.5,1) circle (1.25pt) node[anchor=south] {5};
\filldraw [black] (3,0) circle (1.25pt) node[anchor=west] {6};
\end{tikzpicture}
\caption{A branching structure on a general triangulation.}
\label{fig:triangulation}
\end{figure}
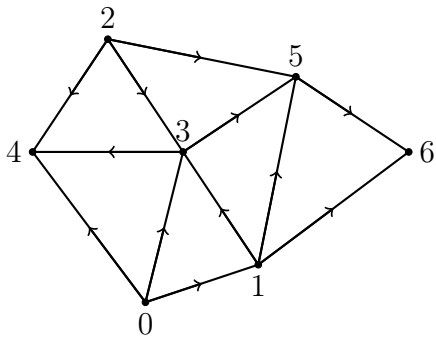

Eq.\ \eqref{eq:Ue definition} means that $U_e$ implements the motion of the magnetic flux across edge $e$ accompanied by the electric charge moving along edges $f_{01}$ of those faces for which $e = f_{12}$. For example, in Figure \ref{fig:triangulation}, we have $U_{35} = X_{35} Z_{23} Z_{13}$, $U_{13} = X_{13} Z_{01}$, and $U_{03} = X_{03}$.

The operators $U_e$ satisfy the same commutation relations as $S_e$, and the vertex relations also agree. One can check that if we impose a Gauss law of the form
\begin{equation}\label{eq:gausstriang}
\prod_{e\supset v} X_e=\prod_{f\in I^0(v)}W_f,
\end{equation}
where $I^0(v)$ is the set of faces such that $v=v_0$ for that face, then $U_e$ satisfy a relation very similar to that of $S_e$:
\begin{equation}
\prod_{e\supset v} U_e=d(v) \prod_{f\supset I^{02}_v} W_f,
\end{equation}
where $d(v)$ is some other $c$-number sign.

It is convenient to regard $d(v)$ and $c(v)$ as 0-chains with values in $\ZZ_2$. It is shown in \cite{BGK} that these chains are homologous, i.e.\ there exists a sign $\eta(e)$ such that
\begin{equation}\label{cycleeq}
\prod_{e\supset v} \eta(e)=c(v)d(v). 
\end{equation}
Thus we obtain the bosonization map
\begin{align}
\begin{split} \label{eq:bosonization}
(-1)^{F_f} = -i \g_f \g^\prime_f & \longleftrightarrow W_f,\\
S_e = i \g_{L(e)} \g^\prime_{R(e)} &\longleftrightarrow \eta(e) U_e.
\end{split}
\end{align}

We note that the sign $\eta(e)$ is not defined uniquely: given one solution to Eq.\ (\ref{cycleeq}), one can get another one by multiplying each $\eta(e)$ by $\zeta(e)$, where $\zeta(e)$ satisfies
\begin{equation}\label{zeta}
\prod_{e\supset v}\zeta(e)=1,\quad\forall v.
\end{equation}
Clearly, there are many solutions of this equation, and consequently the bosonization formulas are not unique. But in a topologically trivial situation, this non-uniqueness is inessential. Indeed, Eq.~(\ref{zeta}) says that $\zeta(e)$ is a 1-cycle, and in a topologically trivial situation every 1-cycle is a boundary of some 2-cycle. That is, every solution of (\ref{zeta}) has the form
\begin{equation}
\zeta(e)=\prod_{f\supset e} \kappa(f)
\end{equation}
for some sign $\kappa(f)$. Such ambiguity is irrelevant in the following sense: the bosonization maps corresponding to $\eta$ and $\eta'=\zeta\eta$ are related by a conjugation by a unitary operator
\begin{equation}
\prod_{\kappa(f)=-1} W_f.
\end{equation}

\section{Examples}

\subsection{Spinless fermion on a square lattice}
As a first example of the bosonization map, consider the theory of complex fermions on a square lattice with nearest-neighbor hopping and an on-site chemical potential $\mu$. The Hamiltonian is
\begin{equation}
H=t\sum_e (c^\dagger_{L(e)} c_{R(e)} + c^\dagger_{R(e)} c_{L(e)}) + \mu \sum_f c^\dagger_f c_f.
\label{eq:H hopping on square}
\end{equation}
To apply our bosonization procedure, we first express \eqref{eq:H hopping on square} in terms of Majorana operators,
\begin{equation}
\begin{split}
H = & \, \frac{t}{2} \sum_e (i \g_{L(e)} \g_{R(e)}^\prime + i \g_{R(e)} \g_{L(e)}^\prime) + \frac{\mu}{2} \sum_f (1+ i \g_f \g_f^\prime)\\
= & \, \frac{t}{2} \sum_e \left(i \g_{L(e)} \g_{R(e)}^\prime - i (\g_{L(e)} \g_{R(e)}^\prime) (-i \g_{L(e)} \g_{L(e)}^\prime)(-i \g_{R(e)} \g_{R(e)}^\prime)\right) \\
&+ \frac{\mu}{2} \sum_f (1+ i \g_f \g_f^\prime).
\end{split}
\label{eq:nearest hopping on square}
\end{equation}
The bosonized Hamiltonian that follows from \eqref{eq:bosonization assumption 1} and \eqref{eq:bosonization assumption 2} is a $\ZZ_2$ gauge theory with Hamiltonian
\begin{equation}
H = \frac{t}{2} \sum_e X_e Z_{r(e)}(1-W_{L(e)}W_{R(e)}) + \frac{\mu}{2} \sum_f (1 - W_f) 
\end{equation}
and a gauge constraint $(\prod_{e \supset v} X_e) W_{\text{NE}(v)}=1$ on each vertex.

\subsection{Spinless fermion on a honeycomb lattice}

\begin{figure}[tb]
\centering
\resizebox{7cm}{!}{%
\begin{tikzpicture}[scale=1]
\draw[thick] (-3.5,0) -- (3.5,0);\draw[thick] (-3.5,1.73205) -- (3.5,1.73205);\draw[thick] (-3.5,-1.73205) -- (3.5,-1.73205);
\draw[thick] (1.5,2.59808) -- (-1.5,-2.59808);\draw[thick] (3.5,2.59808) -- (0.5,-2.59808);\draw[thick] (-0.5,2.59808) -- (-3.5,-2.59808);
\draw[thick] (-1.5,2.59808) -- (1.5,-2.59808);\draw[thick] (-3.5,2.59808) -- (-0.5,-2.59808);\draw[thick] (0.5,2.59808) -- (3.5,-2.59808);
\draw[->][thick] (0,0) -- (1,0);\draw[->][thick] (-2,0) -- (-1,0);
\draw[->][thick] (1,1.73205) -- (2,1.73205);\draw[->][thick] (-1,1.73205) -- (0,1.73205);\draw[->][thick] (-3,1.73205) -- (-2,1.73205);
\draw[->][thick] (1,-1.73205) -- (2,-1.73205);\draw[->][thick] (-1,-1.73205) -- (0,-1.73205);\draw[->][thick] (-3,-1.73205) -- (-2,-1.73205);
\draw[->][thick] (0,0) -- (0.5,0.866025);\draw[->][thick] (2,0) -- (2.5,0.866025);\draw[->][thick] (-2,0) -- (-1.5,0.866025);
\draw[->][thick] (1,-1.73205) -- (1.5,-0.866025);\draw[->][thick] (-1,-1.73205) -- (-0.5,-0.866025);\draw[->][thick] (-3,-1.73205) -- (-2.5,-0.866025);
\draw[->][thick] (-1,1.73205) -- (-0.5,0.866025);\draw[->][thick] (1,1.73205) -- (1.5,0.866025);\draw[->][thick] (-3,1.73205) -- (-2.5,0.866025);
\draw[->][thick] (0,0) -- (0.5,-0.866025);\draw[->][thick] (-2,0) -- (-1.5,-0.866025);\draw[->][thick] (2,0) -- (2.5,-0.866025);
\filldraw [black] (-3,1.73205) circle (1.25pt) node[anchor=north] {1};
\filldraw [black] (-2,0) circle (1.25pt) node[anchor=north] {2};
\filldraw [black] (-1,-1.73205) circle (1.25pt) node[anchor=north] {3};
\filldraw [black] (-1,1.73205) circle (1.25pt) node[anchor=south] {4};
\filldraw [black] (0,0) circle (1.25pt) node[anchor=north]{5};
\filldraw [black] (1,-1.73205) circle (1.25pt) node[anchor=north] {6};
\filldraw [black] (1,1.73205) circle (1.25pt) node[anchor=south] {7};
\filldraw [black] (2,0) circle (1.25pt) node[anchor=north]{8};
\node at (-1,-0.57735){$i$}; 
\node at (-1,0.57735){$j$}; 
\node at (-2,1.1547){$k$}; 
\node at (0,1.1547){$l$}; 
\node at (1,0.57735){$m$}; 
\node at (1,-0.57735){$n$};
\node at (0,-1.1547){$p$}; 
\end{tikzpicture}
}
\caption{A branching structure on triangular lattice.}
\label{fig:triangle}
\end{figure}
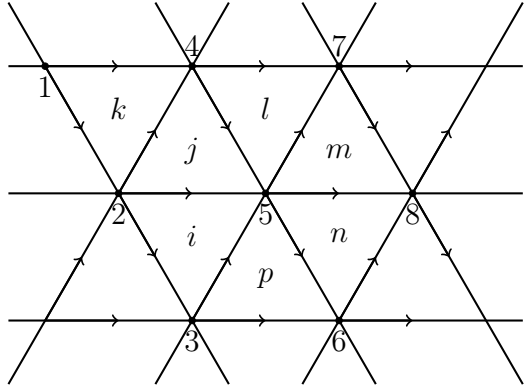

Next, consider fermions living on the faces of a triangular lattice (or on the vertices of a honeycomb  lattice), shown on Fig.~\ref{fig:triangle}. We consider again the nearest neighbor hopping Hamiltonian
\begin{equation}
H=t\sum_e (c^\dagger_{L(e)} c_{R(e)} + c^\dagger_{R(e)} c_{L(e)}) + \mu \sum_f c^\dagger_f c_f.
\label{eq:H hopping on hexagon}
\end{equation}

The hopping operators map as
\begin{equation}
 S_e =  i \g_{L(e)} \g^\prime_{R(e)} \longleftrightarrow \eta(e) U_e
 \label{eq:triangular bosonization map}
\end{equation}
for a suitably chosen sign $\eta(e)$. The sign is chosen so that the vertex relations between $S_e$ and $U_e$ are identical. For the branching structure shown in Fig.~\ref{fig:triangle} one can choose $\eta(e)=1$ for all $e$, so that the bosonization map is simply
\begin{equation}
 S_e \longleftrightarrow U_e.
\end{equation}
With this choice, for the explicitly denoted edges on Fig.~\ref{fig:triangle} the operators $U_e$ defined by \eqref{eq:Ue definition} are
\begin{equation}
\begin{split}
U_{58}&=X_{58},\\
U_{57}&=X_{57}Z_{45},\\
U_{56}&=X_{56}Z_{35},
\end{split}
\end{equation}
and other edges are defined by translation. The bosonized Hamiltonian is
\begin{equation}
H = \frac{t}{2} \sum_e U_e(1-W_{L(e)}W_{R(e)}) + \frac{\mu}{2} \sum_f (1 - W_f)
\label{eq:bosonized nearest neighbor}
\end{equation}
with gauge constraint $(\prod_{e \supset v} X_e)W_{\text{NE}(v)}W_{\text{SE}(v)}=1$ (i.e. 
$(\prod_{e \supset v_5} X_e) W_m W_n = 1$) on each vertex.

It is well known that on the honeycomb lattice the Hamiltonian (\ref{eq:H hopping on hexagon}) gives rise to a dispersion law which has two Dirac points in the Brillouin zone. This is highly non-obvious for the equivalent gauge theory Hamiltonian (\ref{eq:bosonized nearest neighbor}). 

Any fermionic operator with vanishing net fermion parity can be written in terms of $S_e$ and $(-1)^{F_f}$ and thus have a bosonic counterpart. We start from simple examples. To bosonize $c^\dagger_k c_l$ from Fig.~\ref{fig:triangle}, we express it via Majorana operators as
\begin{equation}
\begin{split}
c^\dagger_k c_l =  \frac{1}{4} (\g_k \g_l + \g^\prime_k \g^\prime_l + i \g_k \g^\prime_l + i \g_l \g^\prime_k),
\end{split}
\end{equation}
and then map these Majorana operators in the usual way,
\begin{equation}
\begin{split}
\g_k \g_l & = (i\g_k \g^\prime_j) (i \g_l \g^\prime_j)  \longleftrightarrow   U_{24} U_{45},  \\
\g_k \g^\prime_l &= i(\g_k \g_l) (-i\g_l \g^\prime_l) \longleftrightarrow  i   U_{24} U_{45} W_l,\\
\g_l \g^\prime_k &= (-i) (\g_k \g_l) ( -i\g_k \g^\prime_k)   \longleftrightarrow  -i  U_{24} U_{45} W_k, \\
\g^\prime_k \g^\prime_l &= (-i)(\g_l \g^\prime_k)  ( -i\g_l \g^\prime_l)   \longleftrightarrow -  U_{24} U_{45} W_k W_l.
\end{split}
\end{equation}
This way we obtain
\begin{equation}
\begin{split}
c^\dagger_k c_l = \frac{1}{4} U_{24} U_{45} (1+W_k)(1-W_l).
\label{eq:2nd hopping 1}
\end{split}
\end{equation}

Next, consider the operator $c^\dagger_i c_l = \frac{1}{4} (\g_i \g_l + \g^\prime_i \g^\prime_l + i \g_i \g^\prime_l + i \g_i \g^\prime_l)$. Its first term is
\begin{equation}
\g_i \g_l = (i\g_j \g^\prime_i)  (i \g_l \g^\prime_j) (-i \g_j \g^\prime_j) (-i \g_i \g^\prime_i))\longleftrightarrow U_{25} U_{45} W_j  W_i,
\end{equation}
and the other terms can be computed the same way, giving
\begin{equation}
\begin{split}
c^\dagger_i c_l &= \frac{1}{4} U_{25} U_{45} W_j  W_i (1+W_i)(1-W_l) \\
&=\frac{1}{4} U_{25} U_{45} W_j (1+W_i)(1-W_l).
\label{eq:2nd hopping 2}
\end{split}
\end{equation}

Generalizing from \eqref{eq:2nd hopping 1} and \eqref{eq:2nd hopping 2}, the rule for bosonization of a fermion bilinear $c^\dagger_a c_b$ can be stated as follows. First choose an arbitrary path from face $a$ to face $b$. Start with $(1+W_a)(1-W_b)/4$, and follow the path. When the path passes through a face $f$ by crossing two edges with different orientations, we need to multiply by $W_f$. Then, for each edge $e$ the path crosses, we multiply by $U_e$. For example, following the path $m \rightarrow l \rightarrow k \rightarrow i$, we can write down
\begin{equation}
c^\dagger_i c_m = \frac{1}{4} U_{25} U_{45} U_{57} W_j (1+W_i)(1-W_m).
\label{eq:path1}
\end{equation}
If we use another path $m \rightarrow n \rightarrow p \rightarrow i$, it becomes
\begin{equation}
c^\dagger_i c_m = \frac{1}{4} U_{35} U_{56} U_{58} W_n (1+W_i)(1-W_m).
\label{eq:path}
\end{equation}
The above two formulas only differ by a gauge transformation.


\subsection{The Hubbard model on a square lattice}

The Hamiltonian of Hubbard model (with fermions on faces of a square lattice) is
\begin{equation}
H= t \sum_{e, \, \sigma} (c^\dagger_{L(e),\,\sigma} c_{R(e),\,\sigma} + h.c.) + U \sum_f n_{f \uparrow} n_{f \downarrow}
\end{equation}
where $\sigma\in \{\uparrow,\downarrow\}$ and $n_f = c^\dagger_f c_f$.
It can be viewed as two copies of the nearest neighbor hopping Hamiltonian with an interaction on each face. Similar to \eqref{eq:bosonized nearest neighbor}, the bosonized system is a $\ZZ_2\times\ZZ_2$ gauge theory on the dual lattice with a Hamiltonian
\begin{equation}
H = \frac{t}{2} \sum_{e,\,\sigma} X^\sigma_e Z^\sigma_{r(e)}(1-W^\sigma_{L(e)}W^\sigma_{R(e)}) + \frac{U}{4} \sum_f (1-W^\uparrow_f) (1-W^\downarrow_f)
\end{equation}
with gauge constraints $(\prod_{e \supset v} X^\sigma_e) W^\sigma_{\text{NE}(v)}W^\sigma_{\text{SE}(v)}=1$ for $\sigma=\uparrow,\downarrow$ at each vertex. On each edge, there are two species of spins labeled by $\uparrow$ and $\downarrow$. 

Note that the $SU(2)$ spin symmetry is not manifest in this bosonized description. There exists a version of our bosonization procedure where the $SU(2)$ symmetry is manifest. In that description, one of the $\ZZ_2$ gauge fields is replaced with a bosonic spin which lives on the vertices of the dual lattice. The $SU(2)$ symmetry acts only on this spin variable.

\subsection{Some soluble 2+1D lattice gauge theories}

We have seen a couple of examples where a simple theory of free fermions on a lattice can be rewritten as a rather complicated $\ZZ_2$ lattice gauge theory on the dual lattice. Conversely, one can start with some simple $\ZZ_2$ gauge theory and ask if it can be rewritten as a theory of free fermions. 

The standard 2+1D $\ZZ_2$ lattice gauge theory introduced by F. Wegner \cite{Wegner} can be written in the Hamiltonian form as follows \cite{Kogutreview}. There is a spin on every edge $e$, with Pauli matrices $X_e,Y_e,Z_e$. The physical Hilbert space is a subspace of the tensor product space defined by the Gauss law constraints
\begin{equation}\label{eq:Gaussstandard}
\prod_{e\supset v} X_e=1,\quad \forall v.
\end{equation}
The Hamiltonian is 
\begin{equation}\label{eq:Hstandard}
H=g^2\sum_e X_e+\frac{1}{g^2}\sum_f W_f,
\end{equation}
where $W_f$ is given by Eq.~(\ref{eq:Zf}), as usual. This theory is not integrable and is related by Kramers-Wannier duality to the 3D Ising model. 

To get a $\ZZ_2$ gauge theory which is dual to a fermionic theory, we need to replace Eq.~(\ref{eq:Gaussstandard}) with the modified Gauss law (\ref{eq:gauge constraint at vertex}) on a square lattice, or with (\ref{eq:gausstriang}) on a general triangulation.
The second (potential) term in Eq.~(\ref{eq:Hstandard}) is still gauge-invariant, but the first (kinetic) term is not. To fix this problem we simply replace each $X_e$ with $U_e=X_e Z_{r(e)}$, which is gauge-invariant by construction, and let
\begin{equation}\label{eq:Hsimple}
H'=g^2\sum_e U_e+\frac{1}{g^2}\sum_f W_f.
\end{equation}
Since $W_f$ maps to $-i\gamma_f\gamma'_f$, and $U_e$ maps to $i\gamma_{L(e)}\gamma'_{R(e)}$, the fermionic dual of this gauge theory is a theory of free fermions.

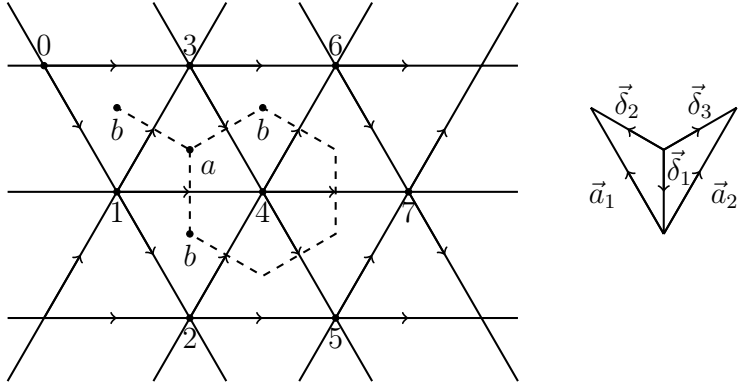
\begin{figure}
\centering
\resizebox{10cm}{!}{%
\begin{tikzpicture}[scale=1]
\draw[thick] (-3.5,0) -- (3.5,0) (-3.5,1.73205) -- (3.5,1.73205) (-3.5,-1.73205) -- (3.5,-1.73205) (1.5,2.59808) -- (-1.5,-2.59808) (3.5,2.59808) -- (0.5,-2.59808)(-0.5,2.59808) -- (-3.5,-2.59808) (-1.5,2.59808) -- (1.5,-2.59808) (-3.5,2.59808) -- (-0.5,-2.59808) (0.5,2.59808) -- (3.5,-2.59808);
\draw[->][thick] (0,0) -- (1,0);\draw[->][thick] (-2,0) -- (-1,0);\draw[->][thick] (1,1.73205) -- (2,1.73205);\draw[->][thick] (-1,1.73205) -- (0,1.73205);\draw[->][thick] (-3,1.73205) -- (-2,1.73205);\draw[->][thick] (1,-1.73205) -- (2,-1.73205);\draw[->][thick] (-1,-1.73205) -- (0,-1.73205);\draw[->][thick] (-3,-1.73205) -- (-2,-1.73205);\draw[->][thick] (0,0) -- (0.5,0.866025);\draw[->][thick] (2,0) -- (2.5,0.866025);\draw[->][thick] (-2,0) -- (-1.5,0.866025);\draw[->][thick] (1,-1.73205) -- (1.5,-0.866025);\draw[->][thick] (-1,-1.73205) -- (-0.5,-0.866025);\draw[->][thick] (-3,-1.73205) -- (-2.5,-0.866025);
\draw[->][thick] (-1,1.73205) -- (-0.5,0.866025);\draw[->][thick] (1,1.73205) -- (1.5,0.866025);\draw[->][thick] (-3,1.73205) -- (-2.5,0.866025);
\draw[->][thick] (0,0) -- (0.5,-0.866025);\draw[->][thick] (-2,0) -- (-1.5,-0.866025);\draw[->][thick] (2,0) -- (2.5,-0.866025);
\draw [dashed, thick] (-1,0.57735) -- (-1,-0.57735) -- (0,-1.1547) -- (1,-0.57735)--(1,0.57735)--(0,1.1547)--(-1,0.57735)--(-2,1.1547);
\draw[thick] (5.5,0.57735) -- (5.5,-0.57735) (5.5,0.57735) -- (6.5,1.1547) (5.5,0.57735)--(4.5,1.1547) (5.5,-0.57735)--(6.5,1.1547) (5.5,-0.57735)--(4.5,1.1547);
\draw[->][thick] (5.5,0.57735) -- (5.5,0);\draw[->][thick] (5.5,0.57735) -- (6,0.866025);\draw[->][thick] (5.5,0.57735) -- (5,0.866025);
\draw[->][thick] (5.5,-0.57735) -- (5,0.288675);\draw[->][thick] (5.5,-0.57735) -- (6,0.288675);
\filldraw [black] (5.40,-0.05) circle (0pt) node[anchor=south west] {$\vec{\delta}_1$};
\filldraw [black] (5,0.866025) circle (0pt) node[anchor=south] {$\vec{\delta}_2$};
\filldraw [black] (6,0.866025) circle (0pt) node[anchor=south] {$\vec{\delta}_3$};
\filldraw [black] (5,0.288675) circle (0pt) node[anchor=north east] {$\vec{a}_1$};
\filldraw [black] (6,0.288675) circle (0pt) node[anchor=north west] {$\vec{a}_2$};
\filldraw [black] (-1,0.57735) circle (1.25pt) node[anchor=north west] {$a$};
\filldraw [black] (0,1.1547) circle (1.25pt) node[anchor=north] {$b$};
\filldraw [black] (-1,-0.57735) circle (1.25pt) node[anchor=north] {$b$};
\filldraw [black] (-2,1.1547) circle (1.25pt) node[anchor=north] {$b$};
\filldraw [black] (-3,1.73205) circle (1.25pt) node[anchor=south] {0};
\filldraw [black] (-1,-1.73205) circle (1.25pt) node[anchor=north] {2};
\filldraw [black] (1,-1.73205) circle (1.25pt) node[anchor=north] {5};
\filldraw [black] (-2,0) circle (1.25pt) node[anchor=north] {1};
\filldraw [black] (0,0) circle (1.25pt) node[anchor=north]{4};
\filldraw [black] (2,0) circle (1.25pt) node[anchor=north]{7};
\filldraw [black] (-1,1.73205) circle (1.25pt) node[anchor=south] {3};
\filldraw [black] (1,1.73205) circle (1.25pt) node[anchor=south] {6};
\end{tikzpicture}
}
\caption{Fermions at the center of faces form a honeycomb lattice. The vectors are defined as $\vec{\delta}_1=(0,-\frac{\sqrt{3}}{3})$, $\vec{\delta}_2=(-\frac{1}{2},\frac{\sqrt{3}}{6})$, $\vec{\delta}_3=(\frac{1}{2},\frac{\sqrt{3}}{6})$ and $\vec{a}_1= \vec{\delta}_2 - \vec{\delta}_1 = (-\frac{1}{2},\frac{\sqrt{3}}{2})$, $\vec{a}_2= \vec{\delta}_3 - \vec{\delta}_1 = (\frac{1}{2},\frac{\sqrt{3}}{2})$. }
\label{fig:triangle2}
\end{figure}

\begin{figure}[h]
\centering
\begin{subfigure}{1\linewidth}
\centering
\includegraphics[width = 8 cm]{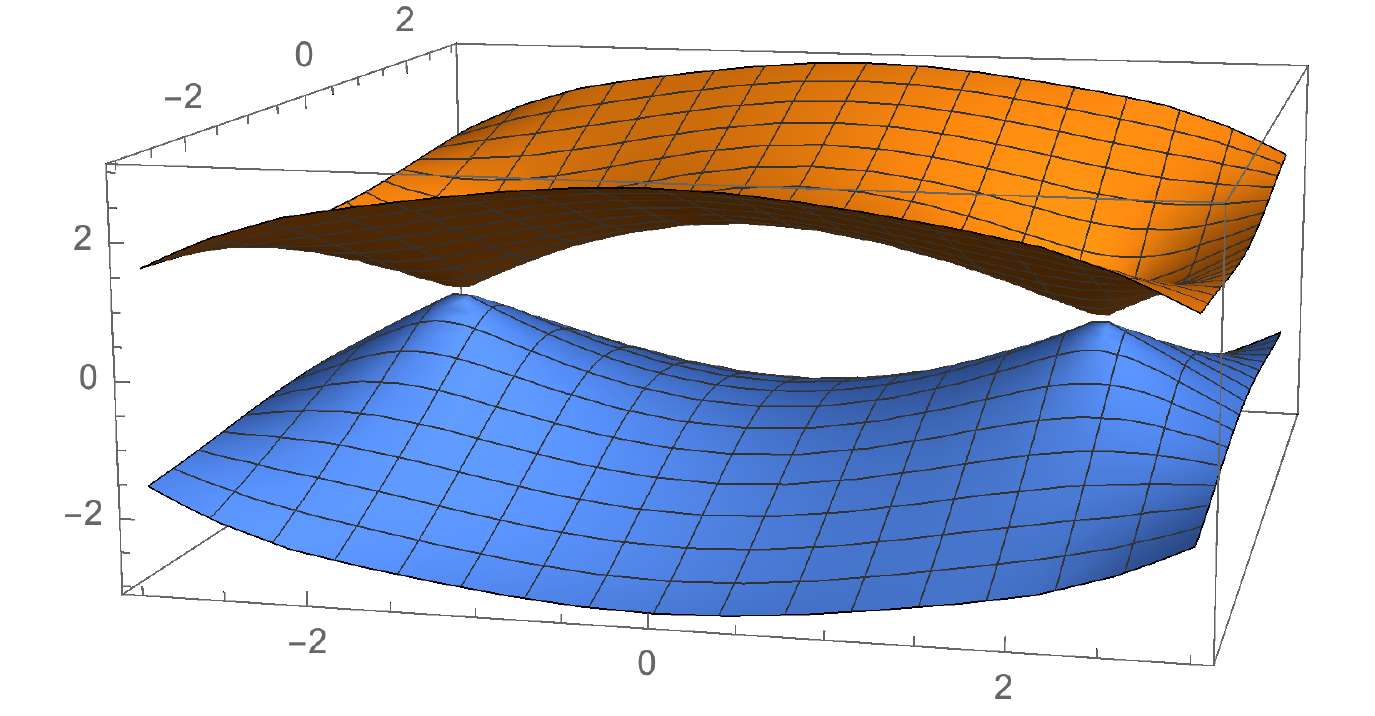}
\caption{For $\mu/t=2$, the band gap closes at $\vec{k}=(\pm \frac{2\pi}{3},0)$, which form two Dirac cones.}
\end{subfigure}
\begin{subfigure}{1\linewidth}
\centering
\includegraphics[width = 7 cm]{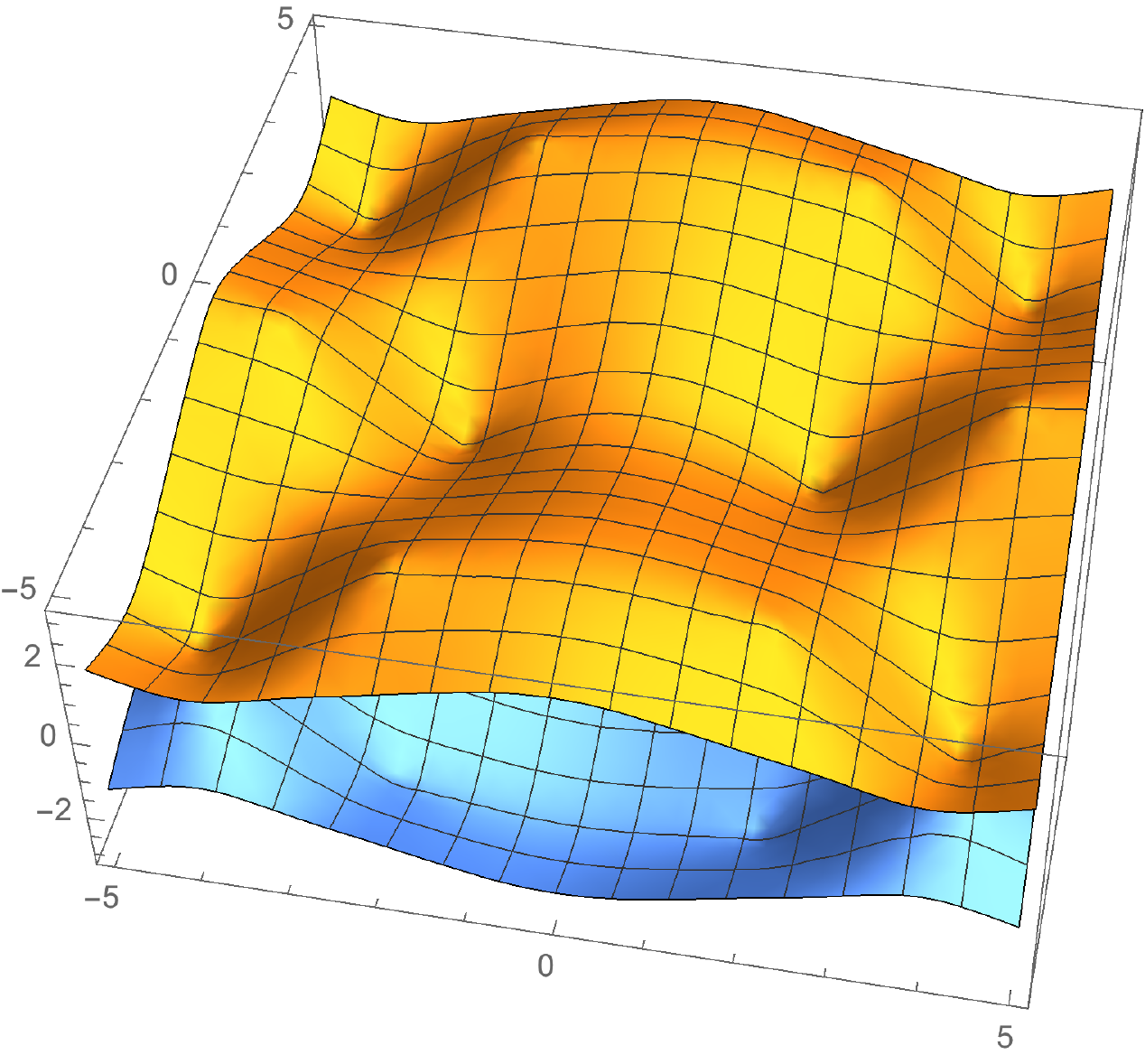}
\caption{Another viewpoint from the top. Two Dirac cones lie in the first Brillouin zone (the hexagon).}
\end{subfigure}  
\caption{(Color online) Band structure of $H_{\_{BdG}}$(equivalent to $H^\prime_f$).}\label{fig:band}
\end{figure}

To analyze this fermionic theory in more detail, let us specialize to the case of a regular triangular lattice (Fig.\ \ref{fig:triangle2}), so that fermions live on the vertices of a regular honeycomb lattice. By bosonization map \eqref{eq:triangular bosonization map}, the Hamiltonian \eqref{eq:Hsimple} is (up to a constant) equivalent to 
\begin{equation}
H'_f = t \sum_e (c_{L(e)} c_{R(e)} - c^\dagger_{L(e)} c^\dagger_{R(e)} + c^\dagger_{L(e)} c_{R(e)} + c^\dagger_{R(e)} c_{L(e)}) + \mu \sum_f  c^\dagger_f c_f,
\end{equation}
where $t=g^2$ and $\mu=2/g^2$. After the usual Fourier transform $c_{\vec{x}} = \frac{1}{\sqrt{N}} \sum_{\vec{k}} e^{i \vec{k} \cdot \vec{x}} c_{\vec{k}}$, the Hamiltonian becomes
\begin{equation*}
H^\prime_f = \sum_{\vec{k}} (\Delta_k \, c_{\vec{k},a} c_{-\vec{k},b} + \mathrm{h.c.}) + \sum_{\vec{k}} (\eps_{\vec{k}} \, c^\dagger_{\vec{k},a} c_{\vec{k},b} + \mathrm{h.c.}) + \mu \sum_{\vec{k}} (c^\dagger_{\vec{k},a}c_{\vec{k},a} + c^\dagger_{\vec{k},b}c_{\vec{k},b})
\end{equation*}
where $\Delta_{\vec{k}}=t(e^{-i \vec{k}\cdot\vec{\delta}_1}-e^{-i \vec{k}\cdot\vec{\delta}_2}-e^{-i \vec{k}\cdot\vec{\delta}_3})$ and $\eps_{\vec{k}}=t(e^{i \vec{k}\cdot\vec{\delta}_1}+e^{i \vec{k}\cdot\vec{\delta}_2}+e^{i \vec{k}\cdot\vec{\delta}_3})$. We can write this using the Bogoliubov-de-Gennes (BdG) formalism as
\begin{equation}
H^\prime_f=\frac{1}{2} \sum_{\vec{k}} \Psi^\dagger_{\vec{k}} H_{\_{BDG}}(\vec{k}) \Psi_{\vec{k}}
\end{equation}
with
\begin{equation}
H_{\_{BDG}}(\vec{k})=
\begin{bmatrix}
\mu & -\Delta^*_{\vec{k}} & \eps_{\vec{k}} & 0 \\
-\Delta_{\vec{k}} & -\mu & 0 & -\eps_{\vec{k}} \\
\eps^*_{\vec{k}} & 0 & \mu & -\Delta^*_{\vec{k}} \\
0 & -\eps^*_{\vec{k}} & -\Delta_{\vec{k}} & -\mu 
\end{bmatrix}
,\quad
\Psi_{\vec{k}}=
\begin{bmatrix}
c_{\vec{k},a} \\ c^\dagger_{-\vec{k},b} \\ c_{\vec{k},b} \\ c^\dagger_{-\vec{k},a}
\end{bmatrix}.
\end{equation}
The eigenvalues are $E(\vec{k})=\pm \sqrt{|\Delta_{\vec{k}}|^2+(|\eps_{\vec{k}}|+\mu)^2}, \, \pm \sqrt{|\Delta_{\vec{k}}|^2+(|\eps_{\vec{k}}|-\mu)^2}$. The gap closes at $k=(\pm \frac{2\pi}{3},0)$ and $\mu/t=2$ ($g=1$). The spectrum is shown in Fig.\ \ref{fig:band}.

\section{Euclidean 3D gauge theories and their fermionic duals}

In this section, we will derive the Euclidean 3D actions for gauge theories dual to some fermionic systems. Before considering  the nearest neighbor hopping Hamiltonian \eqref{eq:nearest hopping on square}, let us first look at the simpler Majorana hopping Hamiltonian on the square lattice,
\begin{equation}
H= -A \sum_e i\g_{L(e)} \g^\prime_{R(e)} - B \sum_f (-i \g_f \g^\prime_f),
\label{eq:majorana hopping H}
\end{equation}
whose bosonic dual is a gauge theory with a Hamiltonian
\begin{equation}
H = - A \sum_e X_e Z_{r(e)} - B \sum_f W_f.
\end{equation}
Without loss of generality, we can assume $A>0$. The Gauss law constraint is
\begin{equation}
G_v  \equiv \left(\prod_{e \supset v} X_e \right) \prod_{e^\prime \subset \text{NE}(v)} Z_{e^\prime} = 1.
\centering
\end{equation}



The partition function is
\begin{equation}
Z=\text{Tr } e^{-\beta H} =  \text{Tr } T^M,
\label{eq:partition function}
\end{equation}
where $T$ is the transfer matrix defined as
\begin{equation}
T= \left(\prod_v \delta_{G'_v, 1 } \right) e^{-\delta \tau H}.
\end{equation}
The prime on $G_v$ means that it acts to bra on the left, which will be clear in later calculations.
The first factor projects to the gauge-invariant sector of the Hilbert space. We can rewrite it using a $\ZZ_2$ Lagrange multiplier field $\lambda_v$ as 
\begin{equation}
 \delta_{G'_v, 1} = \frac{1}{2}(1+G'_v) = \frac{1}{2} \sum_{\lambda_v = \pm 1} (-1)^{\frac{1-\lambda_v}{2} \sum_{e \supset v} \frac{1-X_e}{2}}  \   (-1)^{ \frac{1-\lambda_v}{2} \sum_{e^\prime \subset \text{NE}(v)} \frac{1-Z_{e^\prime}}{2}}.
\label{eq:trick 1}
\end{equation}
Let us define $| m(\tau) \rangle = | \{S_e\} \rangle$ as the configuration of spins (in the $Z_e$ basis). To evaluate the matrix element $\langle m'(\tau + \delta \tau) | T | m(\tau)\rangle$, we insert the ``decomposition of unity'' in terms of a full basis of $X_e$ (momentum) eigenstates, using the identity
\begin{equation}
\langle {S^z}^\prime | f(\sigma^x,\sigma^z) |S^z \rangle = \frac{1}{2} \sum_{S^x = \pm 1} f(S^x,S^z) (-1)^{\frac{1-S^x}{4} \left(2-S^{z\prime} -S^z\right)},
\end{equation}
where we assume $\sigma^x$ is always left to $\sigma^z$ in $ f(\sigma^x,\sigma^z)$. The matrix element is
\begin{equation}
\begin{split}
 & \langle m^\prime(\tau + \delta \tau) | T | m(\tau) \rangle  \\
&\propto \sum_{\{\lambda_v\}}  \left[ \prod_v  (-1)^{\frac{1-\lambda_v}{4} \sum_{e' \subset \text{NE}(v)} (1-S^{z\prime}_{e'})} \  e^{ B \delta \tau \prod_{e' \subset \text{NE}(v)} S^z_{e'}} \right] \times \\
& \qquad  \times \left[ \prod_e \sum_{S^x_e = \pm1} (-1)^{\frac{1-S^x_e}{4} [2-S^{z\prime}_e -S^z_e + \sum_{v^\prime \subset e} (1-\lambda_{v^\prime})]}  \ e^{ A \delta \tau S^x_e S^z_{r (e)}} \right].
\end{split}
\label{eq:Path integral}
\end{equation}
The next order of business is to integrate out the intermediate momentum fields, i.e.~to perform the sum over the $S^x$ in the second bracket. This bracket equals $ \prod_e (e^{A \delta \tau S^z_{r (e)}} + e^{-A \delta \tau S^z_{r (e)}} {S^z_e}^\prime S^z_e \prod_{v^\prime \subset e} \lambda_{v^\prime})$.
To simplify it, we need to consider two cases: $S^z_{r(e)} = 1$ and $S^z_{r(e)} = -1$. First, for $S^z_{r(e)} = 1$, we can simply write
\begin{equation}
\begin{split}
e^{A \delta \tau S^z_{r(e)}} + e^{-A \delta \tau S^z_{r(e)}} {S^z_e}^\prime S^z_e \prod_{v^\prime \subset e} \lambda_{v^\prime} &= 
e^{A \delta \tau} + e^{-A \delta \tau} {S^z_e}^\prime S^z_e \prod_{v^\prime \subset e} \lambda_{v^\prime}  \\
&=C e^{J {S^z_e}^\prime S^z_e \prod_{v^\prime \subset e} \lambda_{v^\prime}},
\end{split}
\label{eq:case1}
\end{equation}
where $C^2 = 2 \sinh (2 A \delta \tau)$ and $\tanh J = e^{-2 A \delta \tau}$. For the other case $S^z_{r(e)} = -1$,
\begin{equation}
\begin{split}
&e^{A \delta \tau S^z_{r(e)}} + e^{-A \delta \tau S^z_{r(e)}} {S^z_e}^\prime S^z_e \prod_{v^\prime \subset e} \lambda_{v^\prime} \\
& \qquad =  C e^{J {S^z_e}^\prime S^z_e \prod_{v^\prime \subset e} \lambda_{v^\prime}} (-1)^{\frac12[2-S^{z\prime}_e -S^z_e + \sum_{v^\prime \subset e} (1-\lambda_{v^\prime})] }.
\end{split}
\label{eq:case2}
\end{equation}

We can combine \eqref{eq:case1} and \eqref{eq:case2} into the single equation
\begin{equation}
\begin{split}
&e^{A \delta \tau S^z_{r (e)}} + e^{-A \delta \tau S^z_{r (e)}} {S^z_e}^\prime S^z_e \prod_{v^\prime \subset e} \lambda_{v^\prime} \\
&\qquad = C e^{J {S^z_e}^\prime S^z_e \prod_{v^\prime \subset e} \lambda_{v^\prime}}  \ (-1)^{ \frac14 [2-S^{z\prime}_e -S^z_e + \sum_{v^\prime \subset e} (1-\lambda_{v^\prime}) ] (1-S^z_{r(e)}) }.
\end{split}
\label{eq:last bracket (simplified)}
\end{equation}
We can now substitute \eqref{eq:last bracket (simplified)} back to \eqref{eq:Path integral} and write the matrix element in the suggestive form
\begin{equation}
\begin{split}
\label{eq:action between layers}
 & \langle m^\prime(\tau + \delta \tau) | T | m(\tau) \rangle  \\
&\quad \propto \sum_{\{\lambda_v\}} \prod_{v, e} e^{ K \prod_{e^\prime \subset \text{NE}(v)} S^z_{e^\prime} +J {S^z_e}^\prime S^z_e \prod_{v^\prime \subset e} \lambda_{v^\prime} } \  (-1)^{\frac12 (1-\lambda_v) \sum_{e' \subset \text{NE}(v)} \frac12 (1-S^{z\prime}_{e'})} \\ 
&\qquad \quad \times (-1)^{\frac12 (1-S^z_{r(e)}) \left[\frac12 (1-S^{z\prime}_e) + \frac12 (1-S^z_e) + \sum_{v^\prime \subset e} \frac12 (1-\lambda_{v^\prime})\right]},
\end{split}
\end{equation}
where $K \equiv B \delta \tau$. 

We can interpret $\lambda_v$ as gauge fields on temporal links. Therefore, the first, exponential term can be thought of as the exponential of the anisotropic Wegner action \cite{Wegner,Kogutreview} 
\begin{equation}\label{eq:Wegneraction}
\sum_f J_f \prod_{e\supset f} S_e,
\end{equation}
where $J_f$ is different for spatial and temporal faces of the 3d lattice.  The rest can be thought of as a topological factor in the partition function which gives the correct anomaly factors of $-1$ for fermionic statistics.  Let $a_i \in C^0(L,\ZZ_2)$ be the $0$-cochain on the $i$-th layer with value $\frac12(1-\lambda_v)$ on vertex $v$. We regard it as the $\ZZ_2$ gauge field on temporal links between the $i$-th and $(i+1)$-th layers. Let $\alpha_i \in C^1(L,\ZZ_2)$ be the $1$-form on  the $i$-th layer that represents the values of $S^z_e$ on the $i$-th layer.  Then we can express the ``topological'' factors in the last line of \eqref{eq:action between layers} as
\begin{align}\notag
&\prod_{v,\,e}  (-1)^{\frac{1-\lambda_v}{2} \sum_{e' \subset \text{NE}(v)} \frac12(1-S^{z\prime}_{e'})} (-1)^{ \frac12(1-S^z_{r(e)})[\frac12(1-S^{z\prime}_e) + \frac12(1-S^z_e) + \sum_{v^\prime \subset e} \frac12(1-\lambda_{v^\prime})] } \\ \label{eq:topological term}
&\quad  \equiv (-1)^{a_{i}(\Delta (\delta \alpha_{i+1}))} (-1)^{\alpha_i ( \Delta (\alpha_{i+1} + \alpha_i + \delta a_i))},
\end{align}
where $\Delta(x)$ is the Poincar\'e dual of $x$ at relative position $(-\frac{1}{2},-\frac{1}{2})$ (i.e.\ $\Delta(\delta_{\text{NE}(v)})=v$ and $\Delta (\delta_e) = r(e)$).
This expression is invariant (up to boundary terms) under the gauge transformation $a_i \rightarrow a_i + f_i + f_{i+1}$ and $\alpha_i \rightarrow \alpha_i + \delta f_i$. 

If we put the Hamiltonian \eqref{eq:majorana hopping H} on a general triangulation instead of a square lattice, its bosonic dual is
\begin{equation}
H = - A \sum_e U_e - B \sum_f W_f.
\end{equation}
Its partition function is
\begin{equation}
Z= \sum_{S^z,\, \lambda} e^{-S_{\text{topo}}} \, e^{ K \sum_{f_s} S^z_{f_s} + J \sum_{f_\tau} S^z_{f_\tau} },
\label{eq:Z majorana}
\end{equation}
where $f_s$ and $f_\tau$ are faces of spatial and temporal types, $S^z_{f} \equiv \prod_{e \subset f} S^z_e$, and
\begin{equation}
e^{-S_{\text{topo}}} = (-1)^{\sum_i [ \int a_i \cup \delta \alpha_{i+1} + \int \alpha_i \cup (\alpha_i + \alpha_{i+1} + \delta a_i) ]}.
\label{eq:Ztopo}
\end{equation}
 Notice that \eqref{eq:Ztopo} is analogous to Chern-Simon action on a general triangulation of 3d manifold
\begin{equation}
S_{\text{CS}} = i \pi \int a \cup \delta a.
\end{equation}
This kind of topological term results in charge-flux attachment and generates fermionic degrees of freedom.

Now, let us go back to the usual fermionic hopping Hamiltonian on a general triangulation
\begin{equation}
H= -2A \sum_e (c^\dagger_{L(e)} c_{R(e)} + c^\dagger_{R(e)} c_{L(e)}) +2B \sum_f c^\dagger_f c_f,
\end{equation}
and its bosonic dual (up to some constant)
\begin{equation}
H = -A \sum_e U_e(1-W_{L(e)}W_{R(e)}) - B \sum_f W_f
\end{equation}
with gauge constraints on vertices $(\prod_{e \supset v} X_e) (\prod_{f} W_f^{\int \delta_v \cup \delta_f} )=1$. The only difference from the Majorana hopping Hamiltonian is the factor $(1-W_{L(e)}W_{R(e)})$. With some careful calculations, one can show the partition function is
\begin{align} \notag
Z=& \sum_{S^z, \, \lambda} e^{-S_{\text{topo}}} \, e^{ K \sum_{f_s} \prod_{e \subset f_s} S^z_{e} + J \sum_{f_\tau} \prod_{e^\prime \subset f_\tau} S^z_{e^\prime} } \\ \notag 
&\times e^{ -\frac{J -\ln 2}{2} \sum_{e_s} [1 + (\prod_{e \subset L(e_s)}S^z_e ) (\prod_{e \subset R(e_s)}S^z_e) ] } \\
& \times e^{-l \sum_{e_s} [1+ (\prod_{e \subset L(e_s)}S^z_e ) (\prod_{e \subset R(e_s)}S^z_e)] (1- {S^z_{e_s}}^\prime S^z_{e_s} \prod_{v \subset e_s} \lambda_{v}) },
\end{align}
where $l$ is taken to be infinity and $e_s$ is a edge on a spatial slice. Taking $l \rightarrow \infty$ imposes additional gauge constraints to the previous lattice gauge theory \eqref{eq:Z majorana}, and the topological term is not affected.

\section{The Fradkin-Shenker model and its fermionic dual}

Another 2d model with a rich structure is the $\ZZ_2$ version of the Abelian Higgs theory. It is a theory of $\ZZ_2$ spins on vertices of a lattice, coupled to a $\ZZ_2$ gauge theory on the lattice edges. This theory was analyzed in detail by Fradkin and Shenker \cite{FS}, and an up-to-date exposition of our current understanding of its phase diagram is given in \cite{FSreview}. 

For simplicity, we work on a square lattice.
We denote by $X_e, Z_e$ the Pauli matrices acting on edge spins and by $S_v^x$ and $S_v^z$ the Pauli matrices acting on the vertex spins. The Hamiltonian of the Fradkin-Shenker model is
\begin{equation}\label{eq:FSHamiltonian}
H = \frac 1{g^2} \sum_f W_f + g^2  \sum_e X_e  + J_p \sum_e Z_e \prod_{v\in e} S_v^z  + J_k \sum_v S_v^x,
\end{equation}
with constraints on vertices
\begin{equation}\label{eq:GaussFS}
 S_v^x \prod_{e \supset v} X_e = 1.
\end{equation}
As usual, $W_f$ is given by (\ref{eq:Zf}). 
From the constraint we see that $S_v^x$ is the electric charge operator for the spin variable, so $S_v^z$ flips the two charge states. The first two terms in Eq.~(\ref{eq:FSHamiltonian}) make up the usual Hamiltonian of the $\ZZ_2$ gauge theory, the last two terms are the potential and kinetic energy of the Ising spin variables. The potential term has been ``covariantized'' by including a factor $Z_e$ to ensure that it commutes with the Gauss law constraints (\ref{eq:GaussFS}). The potential energy term for an edge $e$ can be interpreted as moving the electric charge along $e$, while the kinetic term is simply the total electric charge.

The couplings $J_p$ and $J_k$ are not independent; they should be thought of as monotonic functions of a single parameter, $J$, with different signs of the first derivative w.r.t.~$J$. They are chosen in such a way that the corresponding 3d Euclidean action is invariant under cubic symmetry.  Heuristically, we can think of these couplings as $J_k^{-1} \sim J_p \sim J$, where $J$ is the coupling of the Euclidean Ising theory.

In order to fermionize this theory, we want to introduce hopping operators which move both the electric charge and the gauge field flux. We expect such operators to behave as fermionic hopping operators. A natural candidate is
\begin{equation}\label{eq:fermionicFShop}
V_e =X_e Z_{r(e)} \prod_{v\in r(e)} S_v^z.
\end{equation}
The first two factors are the same as in Section \ref{sec:square}, the last factor is needed to ensure gauge-invariance. If we think of this operator as moving a fermion across $e$, then the fermion number operator must be associated with faces of the lattice. The natural candidate is
\begin{equation}\label{eq:fermionicFSparity}
(-1)^{F_f}=W_f,
\end{equation}
where $W_f$ is given by (\ref{eq:Zf}). It is easy to check that all commutation relations are as expected. 

All gauge-invariant observables can be expressed as functions of $V_e$, $(-1)^{F_f}$, $S_v^x$, and 
\begin{equation}
X^\vee_e=Z_e \prod_{v\in e} S_v^z. 
\end{equation}
The operators $X_e^\vee$ satisfy a constraint on each face:
\begin{equation}\label{eq:dualGaussFS}
\prod_{e\supset f} X^\vee_e=(-1)^{F_f}.
\end{equation}
This looks like a Gauss law for a gauge field on the edges of the dual lattice which is coupled to fermions on the vertices of the dual lattice. The operator $V_{e'}$ anticommutes with $X^\vee_e$ if $e'=e$ and commutes with it otherwise. This is consistent with the identification of $V_e$ as the hopping operator for a fermion. 

It remains to identify the Wilson loop for the dual gauge field in terms of the observables of the Fradkin-Shenker model. On the one hand, we expect the hopping operators for fermions coupled to a gauge field $Z^\vee_e$ to satisfy
\begin{equation}
\prod_{e\supset v} S^F_e=W^\vee_v (-1)^{F_{SW(v)}}(-1)^{F_{NE(v)}}.
\end{equation}
The factor $W^\vee_v$ is the dual Wilson loop: 
\begin{equation}
W^\vee_v=\prod_{e\supset v} Z^\vee_e.
\end{equation}
On the other hand, using the Gauss law (\ref{eq:GaussFS}) we find
\begin{equation}
\prod_{e\supset v} V_e=S_v^x W_{SW(v)}.
\end{equation}
From this we infer
\begin{equation}
W^\vee_v=S_v^x W_{NE(v)}.
\end{equation}
It is easy to see that $W^\vee_v$ anti-commutes with $X^\vee_e$ whenever $v\in e$ and commutes with it otherwise. This matches the expected commutation law for the flux and the electric field. With a little work one can also show that $W^\vee_v$ commutes with $V_e$ for all $v$ and $e$. Finally, it obviously  commutes with $(-1)^{F_f}$ for all $v$ and $f$. Thus all the expected relations are satisfied. 

Note that the Fradkin-Shenker model has two conserved $\ZZ_2$ charges: the net gauge field flux $W=\prod_f W_f$ and the $x$-component of the spin $S^x=\prod_v S_v^x$ (i.e. the net electric charge). After fermionization they correspond to the net  fermion parity $\prod_f (-1)^{F_f}$ and the flux of the dual gauge field $\prod_v W^\vee_v$ times the fermion parity $\prod_f (-1)^{F_f}$, respectively. Local observables are neutral with respect to all these symmetries.

The Hamiltonian (\ref{eq:FSHamiltonian}) can be expressed in terms of fermionic variables as follows:
\bel{
  H = \frac1{g^2} \sum_f (-1)^{F_f} + g^2 \sum_e V_e X_{r(e)}^\vee + J_p \sum_e  X^\vee_e + J_k \sum_v W^\vee_v (-1)^{F_{NE(v)}}.
}
We see that the dual gauge field has a standard kinetic term which moves flux, but an unusual potential term modulated by the fermion parity. The fermion has a rather standard ``potential'' term (essentially, a nonzero chemical potential which for $g^2>0$ favors states with a negative fermion parity), but an unusual kinetic term: it moves both the fermion and the gauge field flux. Thus fluxes are free to move, while the fermion can move only together with a fluxon.

In the limit $g\ra 0$ the gauge field ``freezes out'' and the Fradkin-Shenker model becomes equivalent to the Ising model. In terms of fermionic variables, the fermion parity is frozen to the value $-1$ on all faces. Then we are left with the gauge field on the dual lattice with the standard kinetic and potential terms. This model is related by Kramers-Wannier duality to the Ising model. Another way to get the Ising model from the Fradkin-Shenker model is to take the limit $|J_k|\ra\infty$. In this limit the spins are frozen to be in the eigenstates of $S_v^x$ (with eigenvalue $+1$ or $-1$ depending on the sign of $J_k$). In terms of fermionic variables, in this limit each flux of the dual gauge field is bound to a fermion. The third term in the Hamiltonian can be dropped, while the first two describe an unusual gauge theory coupled to charged fermions which track fluxes. Since the bound state of an electrically charged fermion and a flux is a boson, it is not suprising that this model is equivalent to the Ising model. 

\section{Concluding remarks}

We have shown that an arbitrary 2d fermionic system with a local Hamiltonian is equivalent (in a topologically trivial situation) to a bosonic $\ZZ_2$ gauge theory with a local Hamiltonian. This bosonic gauge theory has a modified Gauss law. The conceptual reason for this is that the usual Gauss law leads to a global 1-form $\ZZ_2$ symmetry which does not suffer from 't Hooft anomaly, while the bosonic dual of a fermionic system must have a global 1-form $\ZZ_2$ symmetry with a nontrivial 't Hooft anomaly \cite{GaiottoAK,BGK}.

We showed that gauge theories with a modified Gauss law can be described by a Euclidean action with a Chern-Simons-like term. This again can be explained in terms of global 1-form symmetries: Chern-Simons terms are a natural way to generate 't Hooft anomalies for such symmetries in three space-time dimensions \cite{gensym}. It is remarkable that although ordinary (nontopological) Euclidean $\ZZ_2$ gauge theories are not integrable in three dimensions, a simple addition of a Chern-Simons-like term makes them equivalent to free fermionic theories and therefore integrable. It would be interesting to understand this from the Euclidean viewpoint and investigate if integrability extends to analogous $\ZZ_N$ gauge theories with $N>2$. 

One may ask if the bosonization prescription  on the  honeycomb lattice introduced by A. Kitaev \cite{Kitaevanyons} is related to ours. In the appendix we show that these two procedures are closely related. Namely, there is a natural way to associate vertices of the honeycomb lattice with edges of the  square lattice. This allows us to rewrite Kitaev's constrained spin system on a honeycomb lattice as a $\ZZ_2$ gauge theory on a square lattice with the modified Gauss law (\ref{eq:gauge constraint at vertex}). Even after this rewriting, Kitaev's bosonization map turns out to be slightly different from ours; in particular, the dual of the fermion parity operator  does not have a simple interpretation in gauge theory terms. The advantage of our procedure is that it works in a straightforward way on an arbitrary lattice and always leads to a $\ZZ_2$ gauge theory with a modified Gauss law. In contrast, a naive generalization of the prescription of \cite{Kitaevanyons} from trivalent lattices to arbitrary 2d lattices leads to a complicated bosonic system, where the bosonic degrees on a vertex $v$ depend on the coordination number of $v$ \cite{Senthil}. 

Another feature of our bosonization procedure is its clear physical interpretation in terms of charge-flux attachment. As a result, it admits a straightforward generalization to higher dimensions. This will be reported elsewhere \cite{YuAnAK}.

\appendix
\section{Relations in the even fermionic algebra}

In this section we sketch a proof that there are no relations between the operators $S_e$ and $(-1)^{F_f}$ besides those described in Section \ref{sec:square} (for the square lattice) and Section \ref{sec:triang} (for a triangulation). The proof applies whenever there are no non-contractible loops in space.

Recall that $\ZZ_2$ operators $S_e$ and $(-1)^{F_f}$ anti-commute if $e\subset f$ and commute otherwise. Also, all operators $(-1)^{F_f}$ commute. The key fact we use is that we can represent all of these operators on the fermionic Fock space. A natural basis in this Fock space is provided by common eigenvectors of $(-1)^{F_f}$. Each such eigenvector is labeled by a 0-chain $\alpha$ with values in $\ZZ_2$ on the dual lattice (i.e. its vertices are faces of the original lattice). If $\alpha(f)=0$, then the state has $(-1)^{F_f}=1$ (i.e.\ is ``empty''). 
If $\alpha(f)=1$, then the state has $(-1)^{F_f}=-1$ (i.e.\ is ``filled'').

Using the commutation relations and the fact that $S_e^2 = 1$, any monomial in $S_e$ and $(-1)^{F_f}$ can be brought to the standard form where all $S_e$ are to the right of all $(-1)^{F_f}$, and all $e$ are distinct. The collection of $e$'s that occur form a 1-chain $\gamma$ with $\ZZ_2$ coefficients (on the dual lattice). On a simply-connected space, any two chains with the same boundary are homologous. The vertex relation for $S_e$ means that, up to a sign, $\prod_{e\subset\gamma} S_e$ depends only on $\partial\gamma$. The sign can be written as a product of operators $(-1)^{F_f}$ for some subset of faces. Note that when such a monomial acts on $\ket{\alpha}$, the result is $\pm \ket{\alpha+\partial\gamma}$.

Suppose now there is a relation in the even fermionic algebra of the form
\begin{equation}
\sum_\gamma C_\gamma \prod_{e\subset\gamma} S_e=0,
\end{equation}
where $C_\gamma$ is some polynomial in the operators $(-1)^{F_f}$. The summation is over a finite set of 1-chains. Using the above remark, we can collect together the terms with the same $\partial\gamma$ and re-write the relation as
\begin{equation}
\sum_{\epsilon} \tilde C_\epsilon \prod_{e\subset \partial^{-1}\epsilon} S_e=0,
\end{equation}
where the summation is over a finite number of 0-chains which are boundaries (i.e. each 0-chain contains an even number of points).

If we let this relation act on $\ket{\alpha}$, we see that the only nonzero $\tilde C_\epsilon$ can occur for $\epsilon=0$. But this means that the relation has the form
\begin{equation}
\tilde C_0=0,
\end{equation}
where $\tilde C_0$ is a polynomial in $(-1)^{F_f}$. Clearly, this can happen only if $\tilde C_0=0$.

\section{Kitaev's honeycomb model as a $\ZZ_2$ gauge theory}

The Hamiltonian of Kitaev's honeycomb model \cite{Kitaevanyons} can be written as
\begin{equation}
H = -J_x \sum_{x-\text{links}} Z^A_j X^B_k -J_y \sum_{y-\text{links}} Y^A_j Y^B_k -J_z \sum_{z-\text{links}} X^A_j Z^B_k.
\label{eq:HKitaev}
\end{equation}
The types of links and lattice structure are shown in Fig. \ref{fig:Kitaev model}. Notice that we have exchanged $X^A \leftrightarrow Z^A$ (with $Y^A \rightarrow -Y^A$) on site $A$ compared to the original model. This way the final bosonization map will become similar to ours. The spin at each site $j$ can be represented by four Majorana operators $b^x_j$, $b^y_j$, $b^z_j$, and $\g_j$. This introduces a redundancy which is eliminated by imposing the constraint $D_j=1$ on each site $j$, where $D_j=b^x_j b^y_j b^z_j \g_j$. The Pauli matrices at each site $j$ are represented by Majorana operators as follows:
\begin{equation}
X_j=i b^x_j \g_j,\;\; Y_j=i b^y_j \g_j, \;\; Z_j=i b^z_j \g_j,
\end{equation}
or equivalently (after multiplying by $D_j$)
\begin{equation}
X_j = -i b^y_j b^z_j , \;\; Y_j=-i b^z_j b^x_j,  \;\; Z_j=-i b^x_j b^y_j.
\label{eq:spin and majorana}
\end{equation}

\tikzstyle{sb}=[circle,draw,fill=black,
inner sep=0pt,minimum size=1.5mm]
\tikzstyle{sa}=[circle,draw,
inner sep=0pt,minimum size=1.5mm]

\begin{figure}
\centering
\begin{tikzpicture}
\begin{scope}[shift={(10,10)}]
\node at (0,0) (B1) [sb] {};\node[below] at (0,-0.05) {$B$};
\node at (2,0) (B2)[sb] {};\node[below] at (2,-0.05) {$B$};
\node at (1,1.73205) (B3)[sb] {};\node[below] at (1,1.68205) {$B$};
\node at (1,-1.73205) (B4)[sb] {};
\node at (-1,1.73205) (B5)[sb] {};
\node at (3,1.73205) (B6)[sb] {};
\node at (0,1.1547) (A1) [sa] {};\node[above] at (0,1.2047) {$A$};
\node at (2,1.1547) (A2)[sa] {};\node[above] at (2,1.2047) {$A$};
\node at (1,-0.57735) (A3)[sa] {};\node[above] at (1,-0.52735) {$A$};
\node at (3,-0.57735) (A4)[sa] {};
\node at (-1,-0.57735) (A5)[sa] {};
\node at (1,2.88675) (A6)[sa] {};
\draw[thick] (B1) -- (A1); \node[left] at (0,0.57735) {z};
\draw[thick] (B1) -- (A3); \node[above] at (0.55,-0.338675) {y};
\draw[thick] (B1) -- (A5);
\draw[thick] (B2) -- (A2); \node[left] at (2,0.57735) {z};
\draw[thick] (B2) -- (A3); \node[below] at (1.55,-0.238675) {x};
\draw[thick] (B2) -- (A4);
\draw[thick] (B3) -- (A1); \node[below] at (0.55,1.51338) {x};
\draw[thick] (B3) -- (A2); \node[above] at (1.55,1.39338) {y};
\draw[thick] (B3) -- (A6);
\draw[thick] (B4) -- (A3);
\draw[thick] (B5) -- (A1);
\draw[thick] (B6) -- (A2);
\end{scope}
\end{tikzpicture}
\caption{}
\label{fig:Kitaev model}
\end{figure}

\begin{figure}
\centering

\resizebox{8cm}{!}{%
\begin{tikzpicture}[scale=2.5]

\node at (0,0) (B1) [sb] {};\node[below] at (0,-0.05) {$B$};
\draw [black] (0,0) circle (14pt) node[anchor=east] {$\g_6$};
\node at (2,0) (B2)[sb] {};\node[below] at (2,-0.05) {$B$};
\draw [black] (2,0) circle (14pt) node[anchor=west] {$\g_4$};
\node at (1,1.73205) (B3)[sb] {};\node[below] at (1,1.68205) {$B$};
\draw [black] (1,1.73205) circle (14pt) node[anchor=west] {$\g_2$};

\node at (0,0.3849) (B1z) [sb] {};\node[below] at (0,0.3849){$b^z_6$};
\node at (0.33333,-0.19245) (B1y) [sb] {};\node[above] at (0.33333,-0.19245){$b^y_6$};
\node at (-0.33333,-0.19245) (B1x) [sb] {};\node[above] at (-0.33333,-0.19245) {$b^x_6$};
\node at (2,0.3849) (B2z) [sb] {};\node[below] at (2,0.3849){$b^z_4$};
\node at (2.33333,-0.19245) (B2y) [sb] {};\node[above] at (2.33333,-0.19245){$b^y_4$};
\node at (1.66667,-0.19245) (B2x) [sb] {};\node[above] at (1.66667,-0.19245){$b^x_4$};
\node at (1,2.11695) (B3z) [sb] {};\node[below] at (1,2.11695){$b^z_2$};
\node at (1.33333,1.5396) (B3y) [sb] {};\node[above] at (1.33333,1.5396) {$b^y_2$};
\node at (0.66667,1.5396) (B3x) [sb] {};\node[above] at (0.66667,1.5396){$b^x_2$};

\node at (0,1.1547) (A1) [sb] {};\node[above] at (0,1.2047) {$A$};
\draw [black] (0,1.1547)circle (14pt) node[anchor=east] {$\g_1$};
\node at (2,1.1547) (A2)[sb] {};\node[above] at (2,1.2047) {$A$};
\draw [black] (2,1.1547)circle (14pt) node[anchor=west] {$\g_3$};
\node at (1,-0.57735) (A3)[sb] {};\node[above] at (1,-0.52735) {$A$};
\draw [black] (1,-0.57735) circle (14pt) node[anchor=west] {$\g_5$};

\node at (0,0.7698) (A1z) [sb] {}; \node[above] at (0,0.7698){$b^x_1$};
\node at (0.33333,1.34715) (A1x) [sb] {}; \node[below] at (0.33333,1.34715){$b^z_1$};
\node at (-0.33333,1.34715) (A1y) [sb] {}; \node[below] at (-0.33333,1.34715){$b^y_1$};
\node at (2,0.7698) (A2z) [sb] {};\node[above] at (2,0.7698){$b^x_3$};
\node at (2.33333,1.34715) (A2x) [sb] {};\node[below] at (2.33333,1.34715){$b^z_3$};
\node at (1.66667,1.34715) (A2y) [sb] {};\node[below] at (1.66667,1.34715){$b^y_3$};
\node at (1,-0.962251) (A3z) [sb] {};\node[above] at (1,-0.962251){$b^x_5$};
\node at (1.33333,-0.384901) (A3x) [sb] {};\node[below] at (1.33333,-0.384901){$b^z_5$};
\node at (0.66667,-0.384901) (A3y) [sb] {};\node[below] at (0.66667,-0.384901){$b^y_5$};

\draw[->] (2.2,1) -- (2.75,1) node[anchor=west] {spin};
\draw[->] (B2y) -- (2.33333,-0.7) node[anchor=north] {Majorana operator};
\draw[thick] (B1z) -- (A1z); \node[left] at (0,0.57735) {z};
\draw[thick] (B1y) -- (A3y); \node[below] at (0.45,-0.238675) {y};

\draw[thick] (B2z) -- (A2z); \node[right] at (2,0.57735) {z};
\draw[thick] (B2x) -- (A3x); \node[below] at (1.55,-0.248675) {x};

\draw[thick] (B3x) -- (A1x); \node[above] at (0.45,1.41338) {x};
\draw[thick] (B3y) -- (A2y); \node[above] at (1.55,1.39338) {y};
\end{tikzpicture}
}
\caption{}
\label{fig:spin to majorana}
\end{figure}
According to \cite{Kitaevanyons}, the fermionic system with a Hamiltonian
\begin{equation}
H= \frac{i}{2} \sum_{e_{jk}} J_{\alpha_{jk}} \g^A_j \g^B_k
\end{equation}
is equivalent to a sector of the system \eqref{eq:HKitaev}. (The index $\alpha$ takes values $x$, $y$, or $z$ depending on the direction of link $jk$.) The sector is described by constraints on each face
\begin{equation}
(i b^z_1 b^x_2) (i b^y_3 b^y_2) (i b^x_3 b^z_4) (i b^z_5 b^x_4) (i b^y_5 b^y_6) (i b^x_1 b^z_6) = 1,
\end{equation}
shown in Fig.\ \ref{fig:spin to majorana}. By Eq.~\eqref{eq:spin and majorana}, the above constraint can be written as
\begin{equation}
(X_1 X_4 X_5 X_6) (Z_1 Z_2 Z_3 Z_4) = 1.
\end{equation}
Further, it is straightforward to define a 1-1 mapping from vertices of the honeycomb lattice to the edges of the square lattice (see Fig.\ \ref{fig:square3}). The above constraint then becomes
\begin{equation}
W_{\_{NE}(v)} \prod_{e \supset v} X_e = 1,
\end{equation}
which agrees with Eq.~(\ref{eq:gauge constraint at vertex}).

\begin{figure}
\centering
\resizebox{6cm}{!}{%
\begin{tikzpicture}[scale=1]
\draw[thick] (-3,0) -- (3,0);\draw[thick] (-3,-2) -- (3,-2);\draw[thick] (-3,2) -- (3,2);
\draw[thick] (0,-3) -- (0,3);\draw[thick] (-2,-3) -- (-2,3);\draw[thick] (2,-3) -- (2,3);
\draw[->] [thick](0,0) -- (1,0);\draw[->][thick] (0,2) -- (1,2);\draw[->][thick] (0,-2) -- (1,-2);
\draw[->][thick] (0,0) -- (0,1);\draw[->][thick] (2,0) -- (2,1);\draw[->][thick](-2,0) -- (-2,1);
\draw[->][thick] (-2,0) -- (-1,0);\draw[->][thick] (-2,2) -- (-1,2);\draw[->][thick](-2,-2) -- (-1,-2);
\draw[->][thick] (-2,-2) -- (-2,-1);\draw[->][thick] (0,-2) -- (0,-1);\draw[->] [thick](2,-2) -- (2,-1);
\filldraw [black] (0,0) circle (1.5pt) node[anchor=north east] {$v$};
\filldraw [black] (1,0) circle (2pt);\filldraw [black] (-1,0) circle (2pt);\filldraw [black] (0,1) circle (2pt);\filldraw [black] (0,-1) circle (2pt);\filldraw [black] (2,1) circle (2pt);\filldraw [black] (1,2) circle (2pt);
\node[left] at (0,1) {1};
\node[above] at (1,2) {2};
\node[right] at (2,1) {3};
\node[below] at (1,0) {4};
\node[right] at (0,-1) {5};
\node[above] at (-1,0) {6};
\draw (1,1) node{NE$(v)$};
\draw [dashed, thick] (-2,-1) -- (2,3);\draw [dashed, thick] (-1,-2) -- (3,2);\draw [dashed, thick] (-3,0) -- (0,3);
\draw [dashed, thick] (2,1) -- (1,2);\draw [dashed, thick] (-1,0) -- (0,-1);\draw [dashed, thick] (0,1) -- (-1,2);\draw [dashed, thick] (-2,-1) -- (-3,0);
\end{tikzpicture}
}
\caption{}
\label{fig:square3}
\end{figure}
To obtain the bosonization map, we combine two Majorana fermions on each $y$-link into one complex fermion at each face of the square lattice. Kitaev's bosonization map then can be phrased as follows:
\begin{enumerate}

\item
Define $Y_f \equiv \prod_{e \in \mathrm{TR}(f)} Y_e$ with $\mathrm{TR} (f)$ being the edges on the top or right of the square $f$.
We identify the fermionic states $ |n_f=0 \rangle$ and $ |n_f=1 \rangle$  with $Y_f=1$ and $Y_f=-1$ states respectively.
\begin{equation}
-i \g_f \g^\prime_f \longleftrightarrow Y_f,
 \label{eq:Kitaev bosonization assumption 1}
\end{equation}

\item
$S_e = i \g_{L(e)} \g^\prime_{R(e)}$ is identified with $U_{r^{-1}(e)}$,
\begin{equation}
 S_e \longleftrightarrow U_{r^{-1}(e)} = Z_e X_{r^{-1}(e)}.
 \label{eq:Kitaev bosonization assumption 2}
\end{equation}
\end{enumerate}
The resulting bosonization map is different from \eqref{eq:bosonization assumption 1} and \eqref{eq:bosonization assumption 2}. The Gauss law and the mapping of hopping operators $S_e$ are the same but identification of the fermion parity operator on the gauge theory side is different.

\end{document}